\documentclass[twocolumn, preprintnumbers,prb,aps,amssymb,showpacs,superscriptaddress]{revtex4}
\usepackage{graphicx}
\usepackage{dcolumn}
\usepackage{wasysym}
\usepackage{bm} \usepackage{color}
\begin{document}

\newcommand{\KBa}{Ba$_{1-x}$K$_{x}$Fe$_2$As$_2$}
\newcommand{\TN}{$T_{\rm N}$}

\newcommand{\ie}{{\it i.e.}}
\newcommand{\eg}{{\it e.g.}}
\newcommand{\etal}{{\it et al.}}
\newcommand{\K}{Ba$_{1-x}$K$_x$Fe$_2$As$_2$}
\newcommand{\KFeAs}{KFe$_2$As$_2$}
\newcommand{\Co}{Ba(Fe$_{1-x}$Co$_x$)$_2$As$_2$}
\newcommand{\Kzero}{$\kappa_0/T$}
\newcommand{\Tc}{$T_c \cong$}
\newcommand{\units}{$\mu \text{W}/\text{K}^2\text{cm}$}
\newcommand{\p}[1]{\left( #1 \right)}
\newcommand{\Dd}[2]{\frac{\text{d} #1}{\text{d}#2}}


\title{
New Phase Induced by Pressure in the Iron-Arsenide Superconductor \KBa
}


\author{E.~Hassinger}
\email{elena.hassinger@usherbrooke.ca}
\affiliation{D\'epartement de physique \& RQMP, Universit\'e de Sherbrooke, Sherbrooke, Qu\'ebec, Canada J1K 2R1}

\author{G.~Gredat}
\affiliation{D\'epartement de physique \& RQMP, Universit\'e de Sherbrooke, Sherbrooke, Qu\'ebec, Canada J1K 2R1}

\author{F.~Valade}
\affiliation{D\'epartement de physique \& RQMP, Universit\'e de Sherbrooke, Sherbrooke, Qu\'ebec, Canada J1K 2R1}

\author{S.~Ren\'e~de~Cotret} 
\affiliation{D\'epartement de physique \& RQMP, Universit\'e de Sherbrooke, Sherbrooke, Qu\'ebec, Canada J1K 2R1}

\author{A.~Juneau-Fecteau} 
\affiliation{D\'epartement de physique \& RQMP, Universit\'e de Sherbrooke, Sherbrooke, Qu\'ebec, Canada J1K 2R1}

\author{J.-Ph.~Reid}
\affiliation{D\'epartement de physique \& RQMP, Universit\'e de Sherbrooke, Sherbrooke, Qu\'ebec, Canada J1K 2R1}

\author{H.~Kim}
\affiliation{Ames Laboratory, Ames, Iowa 50011, USA}

\author{M.~A.~Tanatar}
\affiliation{Ames Laboratory, Ames, Iowa 50011, USA}

\author{R.~Prozorov}
\affiliation{Ames Laboratory, Ames, Iowa 50011, USA} 
\affiliation{Department of Physics and Astronomy, Iowa State University, Ames, Iowa 50011, USA }

\author{B.~Shen}
\affiliation{Center for Superconducting Physics and Materials, National Laboratory of Solid State Microstructures
and Department of Physics, Nanjing University, Nanjing 210093, China}

\author{H.-H.~Wen}
\affiliation{Center for Superconducting Physics and Materials, National Laboratory of Solid State Microstructures
and Department of Physics, Nanjing University, Nanjing 210093, China}
\affiliation{Canadian Institute for Advanced Research, Toronto, Ontario, Canada M5G 1Z8}

\author{N.~Doiron-Leyraud} 
\affiliation{D\'epartement de physique \& RQMP, Universit\'e de Sherbrooke, Sherbrooke, Qu\'ebec, Canada J1K 2R1}

\author{Louis Taillefer}
\email{louis.taillefer@usherbrooke.ca}
\affiliation{D\'epartement de physique \& RQMP, Universit\'e de Sherbrooke, Sherbrooke, Qu\'ebec, Canada J1K 2R1}
\affiliation{Canadian Institute for Advanced Research, Toronto, Ontario, Canada M5G 1Z8}

\date{\today}


\begin{abstract}

The electrical resistivity $\rho$ of the iron-arsenide superconductor~\KBa~was 
measured in applied pressures up to 2.6~GPa for four underdoped samples, with $x \simeq 0.16$, 0.18, 0.19 and 0.21. 
The antiferromagnetic ordering temperature \TN, detected as a sharp anomaly in $\rho(T)$, 
decreases linearly with pressure.
At pressures above $P \simeq 1.0$~GPa, a second sharp
anomaly is detected at a lower
temperature $T_0$, which rises with pressure.
We attribute this second anomaly to the onset of a phase 
that causes a reconstruction of the Fermi surface.
This new phase expands with increasing $x$
and it competes with superconductivity.
%
We discuss the  possibility that a second spin-density wave orders at $T_0$, with a ${\bf Q}$ vector distinct from
that of the spin-density wave that sets in at \TN.


\end{abstract}

\pacs{74.25.Fy, 74.70.Dd}

\maketitle


Superconductivity often appears on the border of antiferromagnetic order,\citep{Monthoux_Nature_2007} 
as in organic conductors, \citep{NDL_PRB_2009} heavy-fermion compounds, \citep{Knebel_review_2009} 
and electron-doped cuprates.\citep{Jin_Nature_2011}
Tuning the system with applied pressure or chemical substitution causes the antiferromagnetic ordering temperature \TN~to
fall and a superconducting phase to eventually appear, with the superconducting transition temperature $T_c$ rising until
the quantum critical point where \TN~goes to zero, and falling thereafter to form a dome-like region of superconductivity
in the phase diagram.
In cuprates, hole doping has the additional effect of inducing the onset of a second phase, with stripe order~\citep{NDL_PhysicaC_2012} -- 
a unidirectional modulation of the spin and charge densities. 
This stripe order competes with superconductivity, and so causes a dip in $T_c$ where it peaks.
Antiferromagnetism and stripe order cause a reconstruction of the Fermi surface, detected for example
in measurements of quantum oscillations and transport properties ({\it e.g.} resistivity, Hall and Seebeck coefficients).\citep{Taillefer_JPCM_2009,Taillefer_ARCMP_2010}

In the iron arsenide BaFe$_2$As$_2$, substitution of K for Ba, Co or Ru for Fe, and P for As
all produce the same type of phase diagram, whereby \TN~falls and a $T_c$ dome surrounds the quantum critical point 
where \TN~$\to 0$.\citep{Canfield_ARCMP_2010}
The application of pressure to BaFe$_2$As$_2$ produces a similar phase diagram.\citep{kim_combined_2011}
The antiferromagnetic order is unidirectional, with wavevector ${\bf Q} = (\pi, 0)$ (or  ${\bf Q} = (0, \pi)$).
It causes the lattice to undergo a transition from tetragonal at high temperature to orthorhombic at low temperature.
The structural transition is either simultaneous with \TN~or slightly before it, as in K-doped~\citep{Avci2012} 
or Co-doped BaFe$_2$As$_2$, \citep{Pratt_PRL_2009} respectively. 
In Co-doped and K-doped BaFe$_2$As$_2$, Fermi-surface reconstruction causes a distinct change in the electrical resistivity $\rho(T)$ below \TN.\citep{Canfield_ARCMP_2010,Shen_PRB_2011}
%
%

\textcolor{white}{.}

\textcolor{white}{.}

In this Article, we report a study of the pressure-temperature phase diagram of \KBa~at  
K concentrations ranging from $x \simeq 0.16$ to $x \simeq 0.21$. 
We find that a pressure in excess of $\sim 1.0$~GPa induces the onset of a new phase
whose extent in the phase diagram increases with $x$, and
whose emergence causes a suppression of superconductivity.
While the underlying order has yet to be determined, 
we discuss the possibility of two successive spin-density waves, the first setting in at \TN~and the second at $T_0$,
with ${\bf Q} = (\pi, 0)$  and ${\bf Q} = (0, \pi)$ (or vice-versa), respectively.


\begin{figure*}[]
\centering
\includegraphics[width=17.6cm]{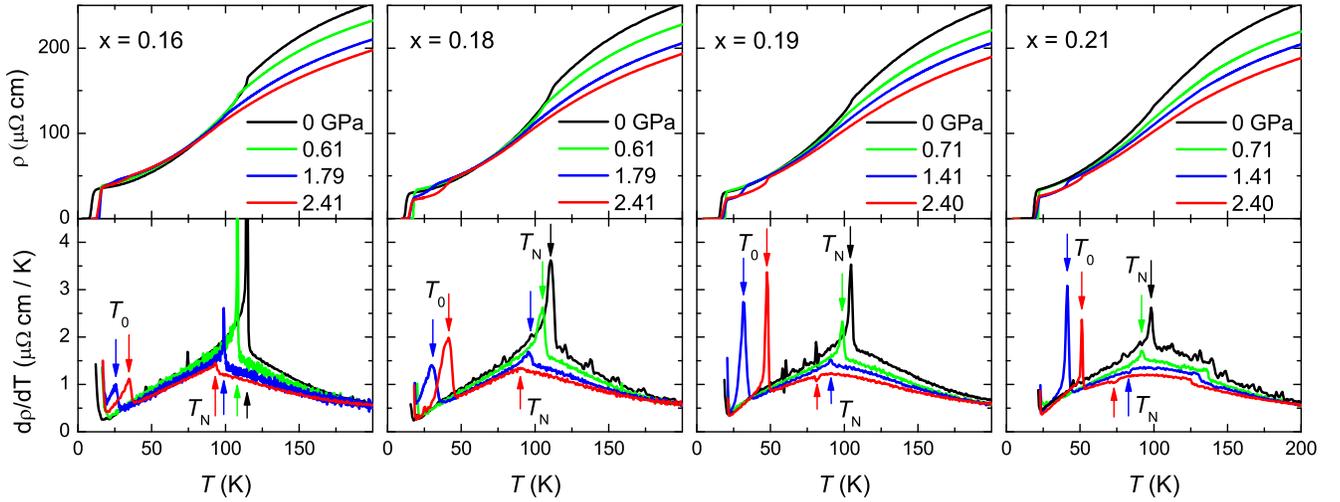}
\caption{
{\it Top}:
In-plane electrical resistivity of \KBa~with $x = 0.16$, $x = 0.18$, $x = 0.19$ and $x = 0.21$ (different columns) for different pressures, as indicated.
{\it Bottom}: 
Temperature derivative of the data in the top panels.
The peak (dip) near 100~K signals the onset of antiferromagnetic order at \TN.
The peak at lower temperature, seen for $P > 1.0$~GPa, signals the onset of a second phase at $T_0$.
}
\label{300K}
\end{figure*}



\begin{figure*}[]
\centering
\includegraphics[width=17.6cm]{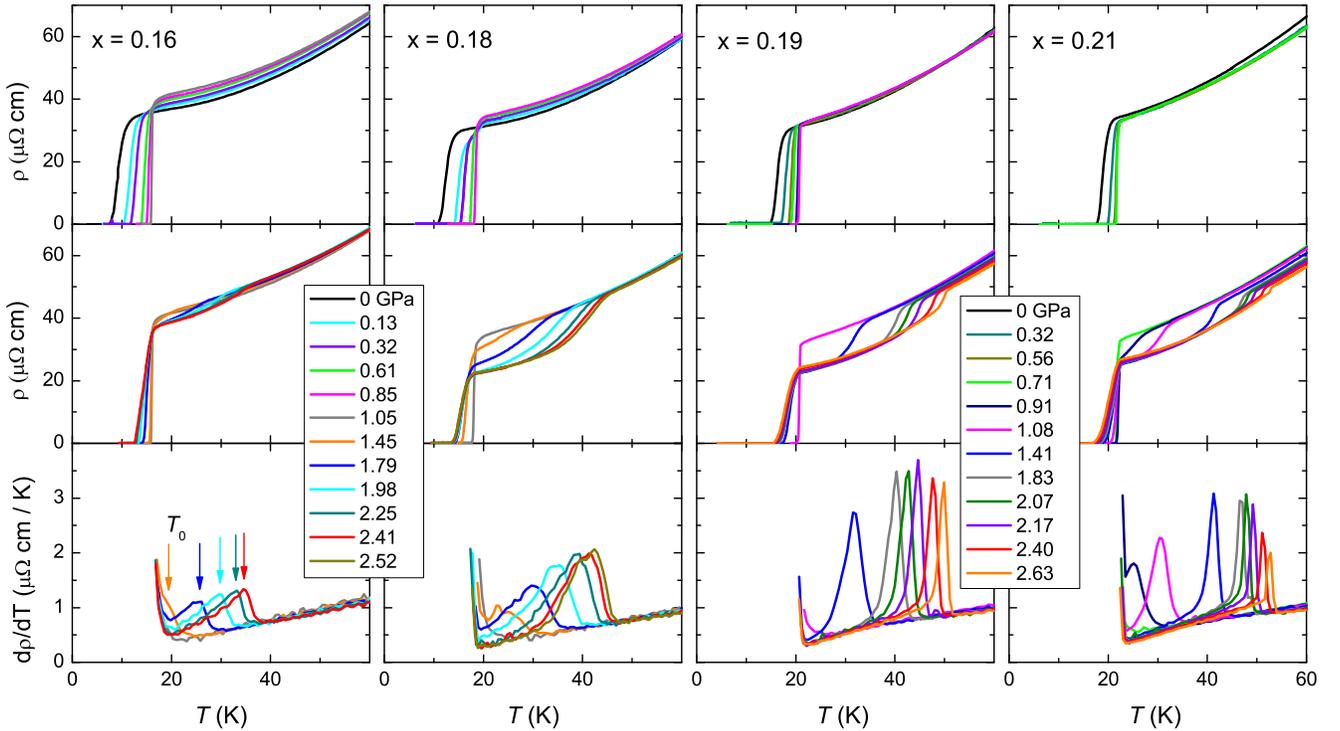}
\caption{
{\it Top and middle}:
Resistivity of \KBa~with $x = 0.16$, $x = 0.18$, $x = 0.19$ and $x = 0.21$ (different columns) below 60~K, for pressures as indicated.
%
{\it Bottom}: 
Temperature derivative of the curves in the middle panel.
The arrows mark the anomaly at $T_0$.
}
\label{Tc}
\end{figure*}



{\it Methods.--}
Single crystals of \KBa~were grown from self flux.\citep{Growth}
Four underdoped samples were measured, with a superconducting transition temperature 
$T_c = 7.3 \pm 0.5$~K, $10.5 \pm 0.5$~K, $15.0 \pm 0.5$~K and $18.0 \pm 0.5$~K, respectively.
Using the relation between $T_c$ and the nominal K concentration $x$ reported in ref.~\onlinecite{Avci2012},
we obtain $x = 0.161$, 0.175, 0.194 and 0.207, respectively.
For simplicity, we label these  $x = 0.16$, 0.18, 0.19 and 0.21.
These $x$ values are also consistent with the measured
antiferromagnetic ordering temperature \TN~(which 
coincides with the structural transition from tetragonal to orthorhombic),~\citep{Avci2012}
equal to 
$115 \pm 1$~K, $111 \pm 1$~K, $105 \pm 1$~K and $98 \pm 1$~K, respectively.
%
%
Hydrostatic pressures up to 
2.63~GPa were applied with a hybrid piston-cylinder cell,\citep{walker_nonmagnetic_1999}
using a 50:50 mixture of n-pentane:isopentane.\citep{duncan_high_2010}
The pressure was measured via the superconducting transition of a lead wire inside the pressure cell. 
The electrical resistivity $\rho$ was measured for a current in the basal plane of the orthorhombic crystal structure,
with a standard four-point technique using a Lakeshore ac-resistance bridge. 
%
%
When a magnetic field was applied, it was along the $c$ axis, normal to the basal plane.
The transition temperatures are defined as follows: 
$T_c$ is where $\rho=0$; 
\TN~and $T_0$ are extrema in the derivative $d\rho/dT$ (Fig.~\ref{300K}).
%


{\it Resistivity.--}
In Fig.~\ref{300K}, the resistivity of \KBa~is plotted as a function of temperature, at four representative pressures for $x=0.16$, 0.18, 0.19 and 0.21.
To remove uncertainties coming from the geometric factors of the different samples, we set $\rho = 300~\mu\Omega$~cm at $T=300$~K,
in agreement with previous studies.\citep{Shen_PRB_2011}
The antiferromagnetic transition at \TN~is detected as a sharp peak in the derivative $d\rho/dT$, which becomes
less and less pronounced with pressure. 
For $x = 0.19$ and $x = 0.21$, the anomaly changes from a peak to a dip, above $P = 1.83$~GPa and $P = 1.08$~GPa, respectively. 
The same effect is observed with increasing $x$ at ambient pressure.\citep{Shen_PRB_2011}
This change seems to happen when $T_N$ falls below $\sim 87$~K.

Above \TN , $\rho$ decreases with pressure, at the rate of $-10$\%/GPa at 200~K. 
Below \TN , the pressure dependence of $\rho$ has nearly vanished 
and $\rho = \rho_0 + A T^n$, with $n=2$ for $x=0.16$ and 0.18, $n = 1.85 \pm 0.05$ for $x = 0.19$ and 0.21, in agreement with $n = 1.90$ at
$x = 0.20$ reported previously.\cite{Shen_PRB_2011}. 
$n$ is independent of pressure and $A$ decreases only slightly with pressure.
The drop in $\rho(T)$ below \TN~is due to the reconstruction of the Fermi surface caused by the antiferromagnetic order,
where the loss in carrier density is more than compensated by the reduction in scattering, as in stoichiometric BaFe$_2$As$_2$.\citep{Shen_PRB_2011}

For $P > 1$~GPa, a second drop in $\rho(T)$ is observed at lower temperature.
It produces a peak in $d\rho/dT$ similar to that at \TN, revealing the onset of a second Fermi-surface reconstruction,
at a temperature labelled $T_0$.
In Fig.~\ref{Tc}, a zoom at low temperature shows that $T_0$ moves up under pressure, in contrast to \TN~which moves down.

The superconducting transition moves up with pressure initially, and it becomes sharper where $T_c$ is maximal. 
At pressures where the new phase is present, 
$T_c$ moves down with pressure and the transition widens. 
At $ P > 1$~GPa, the onset of the superconducting drop is independent of pressure.


\begin{figure}[]
\centering
\includegraphics[width=8.5cm]{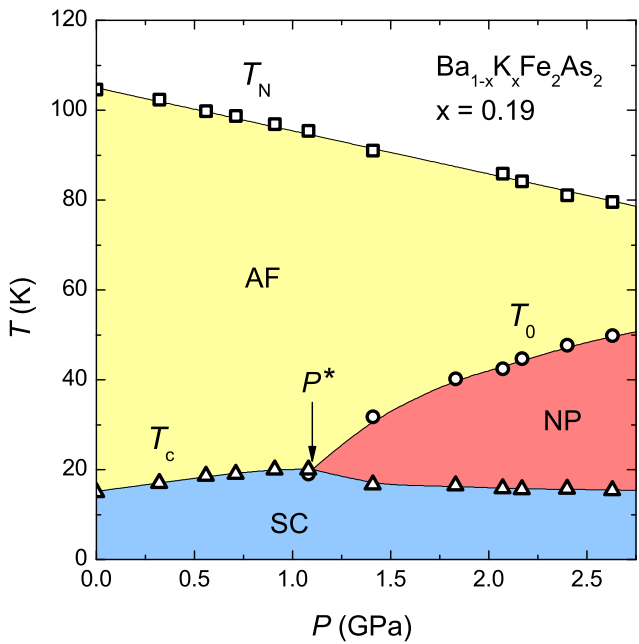}
\caption{
Pressure-temperature phase diagram of Ba$_{1-x}$K$_{x}$Fe$_2$As$_2$ for $x = 0.19$, 
showing the antiferromagnetic (AF) ordering temperature \TN , the superconducting (SC) transition temperature $T_c$
and the anomaly at $T_0$, as a function of pressure.
Error bars are smaller than the size of the symbols.
We attribute the anomaly at $T_0$ to the onset of a new phase (NP), whose order has yet to be determined.
$P^\star$ is the critical pressure above which the new phase is present.
}
\label{phasediag}
\end{figure}



\begin{figure}[]
\centering
\includegraphics[width=8.5cm]{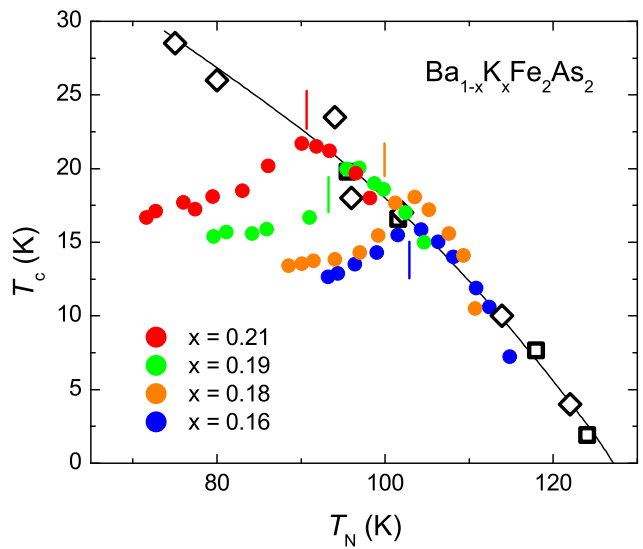}
\caption{
Evolution of the superconducting temperature $T_c$ of underdoped \KBa~
as a function of the corresponding antiferromagnetic temperature \TN, obtained by varying either $x$ at ambient pressure (open black symbols) 
or pressure at fixed $x$, for 4 values of $x$, as indicated. 
%
The data come from neutron (open diamonds, ref.~\onlinecite{Avci2012})
and transport (open squares, ref.~\onlinecite{Shen_PRB_2011}) measurements.
The line is a guide to the eye.
The small vertical lines indicate the position of $P^{\star}$.
Note that pressure and doping have the same effect in both decreasing \TN~and increasing $T_c$, until
pressure induces the onset of a new phase at $P^{\star}$, whereupon $T_c$ drops from its otherwise 
monotonic increase vs $P$ and $x$.
}
\label{TcvsTN}
\end{figure}


\textcolor{white}{.}

\textcolor{white}{.}


{\it Phase diagram.--}
In Fig.~\ref{phasediag}, the evolution of \TN, $T_0$ and $T_c$ with pressure is displayed on a phase diagram for $x=0.19$.
%
%
Initially, $T_c$ rises as \TN~falls, reflecting the competition between antiferromagnetic and superconducting phases.
At low pressure, the pressure-tuned competition mimics the well-known concentration-tuned competition (Fig.~4).
$T_c$ reaches a maximal value of $20.0 \pm 0.2$~K at $P \simeq 1$~GPa, and then it falls.
The peak in $T_c$ coincides with the point where the $T_0$ and $T_c$ lines intersect; we label this  pressure $P^{\star}$. 
(The point $T_0 < T_c$ at 1.08~GPa was determined by the application of a magnetic field to lower $T_c$; see Fig.~5.)
%


\begin{figure}[]
\centering
\includegraphics[width=8.5
cm]{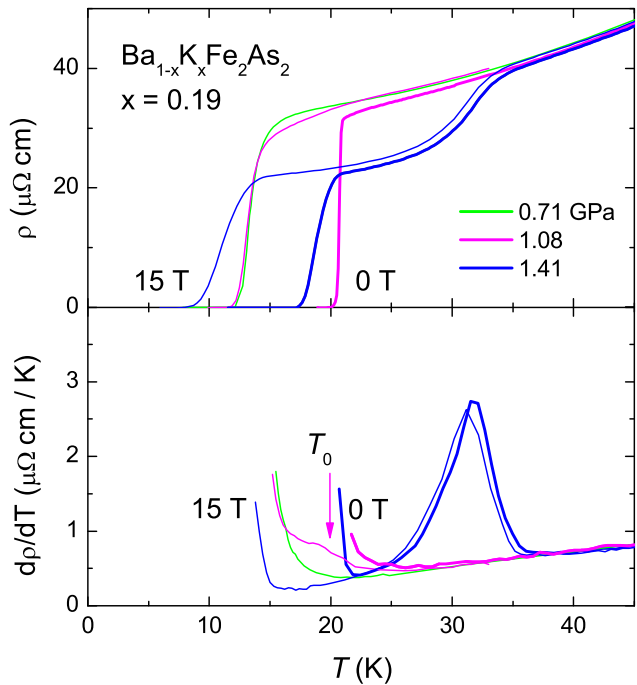}
\caption{
{\it Top}:
Resistivity of sample $x=0.19$ as a function of temperature for pressures as indicated, at $H=0$~T (thick lines) and $H=15$~T (thin lines).
{\it Bottom}:
The temperature derivatives of the curves in the top panel.
Note how the anomaly in the 1.08~GPa curve appears when the superconducting transition is lowered by the magnetic field,
at $T_0$ (arrow).
}
\label{15T19}
\end{figure}

\begin{figure}[]
\centering
\includegraphics[width=8.5cm]{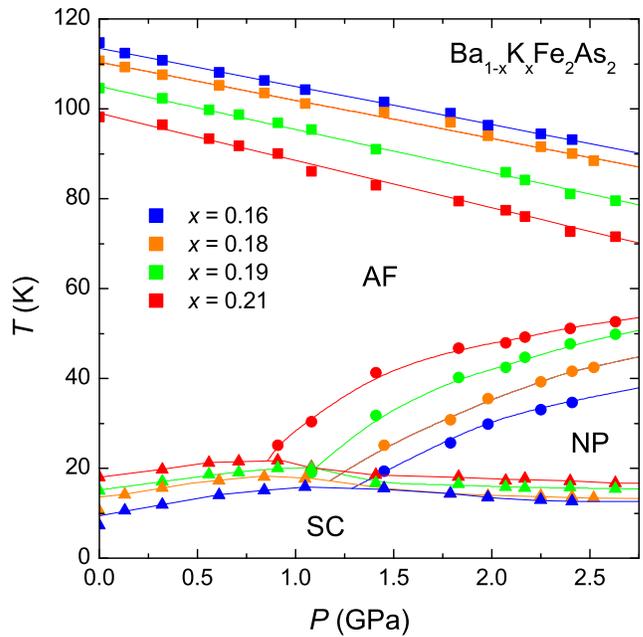}
\caption{
Phase diagram as in Fig.~3, for all four concentrations $x$, as indicated by color-coded symbols.
}
\label{phasediagall}
\end{figure}
\begin{figure}[]
\centering
\includegraphics[width=6cm]{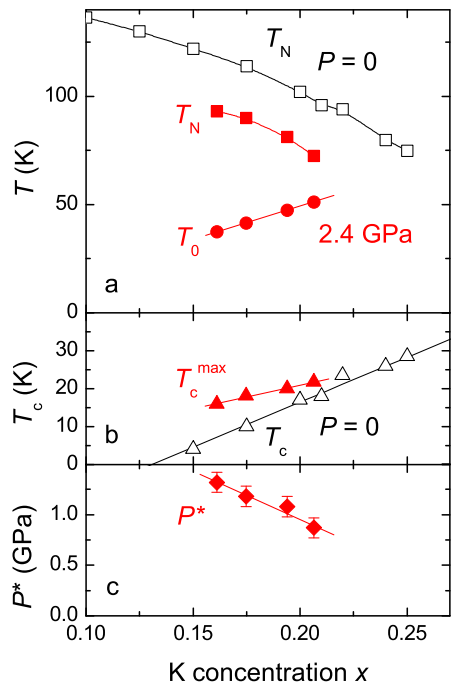}
\caption{ 
Evolution of various properties of Ba$_{1-x}$K$_{x}$Fe$_2$As$_2$ with K concentration $x$.
a) Antiferromagnetic ordering temperature \TN~at $P=0$ (open black squares,\ from neutron scattering data~\citep{Avci2012}) 
and at $P=2.4$~GPa (full red squares, from Fig.~6).  Onset temperature $T_0$ for the new phase, at $P=2.4$~GPa 
(full red circles, from Fig.~6).
All lines are a guide to the eye.
b) $T_c$ at zero pressure (open triangles, from magnetization data\citep{Avci2012}); the linear fit (line) is used to define
the value of $x$ in our samples.
Also shown is the maximal value of $T_c$ attained under pressure, labelled $T_c^{\rm max}$ (full triangles),
with a linear fit (line).
c) The pressure $P^{\star}$ where the new phase appears (diamonds); the line is a linear fit. 
}
\label{dopdep}
\end{figure}

%
Qualitatively identical phase diagrams are obtained for all four samples (Fig.~\ref{phasediagall}). 
With increasing $x$, the antiferromagnetic phase shrinks, while the new phase expands (to higher temperature and lower pressure). 
The peak in $T_c(P)$ correlates with the appearance of the new phase, i.e. it coincides with $P^\star$.
As shown in Fig.~\ref{dopdep}a, $T_N$ decreases with doping the same way at zero pressure and at 2.4 GPa. 
At 2.4~GPa, $T_0$ increases linearly with doping, so that $T_0$ and $T_N$ are expected to become equal
at $x \simeq 0.23$.
The maximum $T_c$ attained under pressure, $T_c^{\rm max}$,  increases with $x$ (Fig.~\ref{dopdep}b);
at high $x$, it approaches the value of $T_c$ at zero pressure since
$P^{\star}$ moves down with $x$ (Fig.~\ref{dopdep}c).

\textcolor{white}{.}


{\it Discussion.--}
A drop in the resistivity could have a number of possible origins.
First, we rule out the possibility of an incomplete superconducting transition by studying the effect of a magnetic field.
In Fig.~\ref{15T19}, $\rho(T)$ for $x = 0.19$ is shown at $H=0$ and $H=15$~T. 
While $T_c$ shifts down by $\sim 7$~K, $T_0$ is only suppressed by about 0.7~K. 
%
A second possibility is a Lifshitz transition.
Within a single antiferromagnetic phase, the Fermi surface can undergo a second reconstruction below the original one at \TN~when
the spin-density-wave order parameter exceeds a certain critical value.
However, such a Lifshitz transition is unlikely to be the explanation here, as $T_0$ and \TN~respond in opposite directions 
to both pressure (Fig.~6) and K concentration (Fig.~7). 
%

%
%
%


Instead, the phenomenology strongly suggests that a second phase transition occurs at $T_0$, to a new phase with currently unknown order. 
Let us mention two possible density-wave scenarios.
The first is a charge-density wave.
ARPES data on BaFe$_2$As$_2$ has revealed highly parallel sections of the Fermi surface inside the antiferromagnetic phase.\citep{Kondo_PRB_2010}
Such features suggest the possibility of an incommensurate charge-density-wave instability favored by the good nesting conditions,
which may be improved by tuning $x$ and applying pressure.

\textcolor{white}{.}

\textcolor{white}{.}

A second possibility is that \TN~and $T_0$ are the onset temperatures of two successive spin-density-wave phases.
The situation is reminiscent of the two successive charge-density-wave transitions in the rare-earth tri-tellurides $R$Te$_3$,\citep{Ru_PRB_2008}
where nesting at a wavevector ${\bf Q}_1$ gaps out part of the Fermi surface below the first transition, at $T_{c1}$, 
and nesting at a wavevector ${\bf Q}_2$, perpendicular to ${\bf Q}_1$, further gaps out 
the Fermi surface below the second transition, at $T_{c2} < T_{c1}$. 
By changing the rare-earth ion $R$ from Dy to Tm, the two transition temperatures go in opposite directions:
$T_{c1}$ drops while $T_{c2}$ rises.\citep{Ru_PRB_2008}
This is interpreted as follows:
as the first gap, $\Delta_1$, decreases, more of the Fermi surface remains after reconstruction below $T_{c1}$ and so more of it can take part in the nesting
at ${\bf Q}_2$, thus producing a stronger gap $\Delta_2$, and hence a larger $T_{c2}$.\citep{Ru_PRB_2008}

The fact that \TN~and $T_0$ go in opposite directions with pressure in \KBa~suggests a similar picture.
The first spin-density wave orders below \TN~with ${\bf Q}_1 = (\pi, 0)$ (within a given orthorhombic domain),
causing two of the four electron pockets in the Fermi surface to reconstruct.
The proposed scenario is that a second spin-density wave orders below $T_0$, with a different wave vector, ${\bf Q}_2$.
It is conceivable that ${\bf Q}_2 \simeq (0, \pi)$, causing the other two electron pockets to reconstruct.


Three other features of our data appear consistent with a scenario of two related spin-density-wave phases.
%
%
First, the two transitions, at \TN~and $T_0$, cause similar changes in the resistivity:
$\rho(T)$ drops in both cases, and the drop is of comparable sharpness (see $d\rho/dT$ in Fig.~\ref{300K}).
Secondly, with increasing doping or pressure, the anomaly in $d\rho/dT$ becomes weaker at $T_N$ but stronger at $T_0$. 
This is consistent with nesting conditions that deteriorate at ${\bf Q}_1$ and improve at ${\bf Q}_2$ with increasing $P$ or $x$.
%
%
Finally, the new phase appears to compete with superconductivity, as does the antiferromagnetic order.
 Below $P^{\star}$, $T_c$ increases while $T_N$ decreases with pressure.
Above $P^{\star}$, as the new phase grows, $T_c$ drops and the dependence of $T_c$ on \TN~deviates 
(Fig.~\ref{TcvsTN}).
%
%


In summary, we report an anomaly in the temperature dependence of the resistivity of underdoped \KBa~for $P > 1$~GPa 
that signals the onset of a Fermi-surface reconstruction at a temperature $T_0$ below the antiferromagnetic temperature \TN.
We attribute this reconstruction to a new phase that onsets below $T_0$.
Whether this phase involves order in the spin, charge or orbital degree of freedom remains to be determined.
However, the overall phenomenology is consistent with a scenario of two related spin-density-wave phases setting
in successively at \TN~and $T_0$, with wavevectors ${\bf Q}_1 \simeq (\pi, 0)$ and ${\bf Q}_2 \simeq (0, \pi)$, respectively.


We thank A.~V.~Chubukov, R.~Fernandes, I.~R.~Fisher, S.~A.~Kivelson, J.~Schmalian and R.~Thomale for fruitful discussions 
and J.~Corbin for his assistance with the experiments. 
The work at Sherbrooke was supported by a Canada Research Chair, CIFAR, NSERC, CFI and FQRNT.
The work at the Ames Laboratory was supported by the DOE-Basic Energy Sciences under Contract No. DE-AC02-07CH11358.
The work in China was supported by NSFC and the MOST of China (\#2011CBA00100).



\begin{references}

\bibitem{Monthoux_Nature_2007}
P.~Monthoux {\it et al.}, 
Nature {\bf 450}, 1177 (2007).

\bibitem{NDL_PRB_2009}
N.~Doiron-Leyraud {\it et al.}, 
Phys. Rev. B {\bf 80}, 214531 (2009).

\bibitem{Knebel_review_2009}
G.~Knebel {\it et al.}, C. R. Phys. {\bf 12}, 542 (2011).

\bibitem{Jin_Nature_2011}
K.~Jin {\it et al.}, 
Nature {\bf 476}, 73 (2011).

\bibitem{NDL_PhysicaC_2012}
N.~Doiron-Leyraud and L.~Taillefer,
Physica C {\bf 481}, 161 (2012).

\bibitem{Taillefer_JPCM_2009}
L.~Taillefer,
J. Phys.: Condens. Matter {\bf 21}, 164212 (2009).

\bibitem{Taillefer_ARCMP_2010}
L.~Taillefer,
Annu. Rev. Condens. Matter Phys. {\bf 1}, 51 (2010).

\bibitem{Canfield_ARCMP_2010}
P.~C.~Canfield and S.~L.~Budk'o, 
Annu. Rev. Condens. Matter Phys. {\bf 1}, 27 (2010).

\bibitem{kim_combined_2011}
S.~K.~Kim {\it et al.},
Phys. Rev. B {\bf 84}, 134525 (2011).

\bibitem{Avci2012}
S.~Avci  {\it et al.}, 
Phys. Rev. B {\bf 85}, 184507 (2012).

\bibitem{Pratt_PRL_2009}
D.~K.~Pratt {\it et al.},
Phys. Rev. Lett. {\bf 103}, 087001 (2009).





\bibitem{Shen_PRB_2011}
B.~Shen {\it et al.},
Phys. Rev. B {\bf 84}, 184512 (2011).

\bibitem{Growth}
H.-Q.~Luo {\it et al.},
Supercond. Sci. Technol. {\bf 21}, 125014 (2008).
 
\bibitem{walker_nonmagnetic_1999}
I.~R.~Walker,
Rev. Sci. Instrum. {\bf 70}, 3402 (1999).

\bibitem{duncan_high_2010}
W.~J.~Duncan {\it et al.},
J. Phys.: Condens. Matter {\bf 22}, 052201 (2010).



\bibitem{Kondo_PRB_2010}
T.~Kondo {\it et al.},
Phys. Rev. B {\bf 81}, 060507 (2010).

\bibitem{Ru_PRB_2008}
N.~Ru {\it et al.},
Phys. Rev. B {\bf 77}, 035114 (2008).



\end{references}
\end{document}